\documentclass{elsart}

\usepackage{amssymb}
\usepackage{amsmath}
\usepackage{epsf,colordvi,color,amsbsy}
\usepackage[dvips]{graphicx}
\begin{document}

\begin{frontmatter}

\title{A general parametric model for the dynamic dark energy}

\author[]{Stefano Sello \corauthref{}}

\corauth[]{stefano.sello@enel.it}

\address{Mathematical and Physical Models, Enel Research, Pisa - Italy}

\begin{abstract}

In the present work we suggest new and more generalized parameterizations for the Equation of State, EoS, of dark energy, maintaining the basic structure of two-parameters CPL-model, but covering both the past and the future of the cosmic history, without divergences and consistently with the current observational data. We propose two generalizations, starting from the extended MZp-model by Ma and Zhang, 2011, the $\xi$MZp-model and the DFp-model. The potential advantages of using these new formulations is their extended range of validity, mainly in the future, to determine possible future scenarios of the cosmic evolution.

\end{abstract}
\end{frontmatter}

\section{Introduction}

Although there is now a strong confidence level on observational evidence for the existence of dark energy ($\sim 5.4 \sigma$),the responsible driver of the recent accelerated expansion of the universe (Riess et al., 1998 and Perlmutter et al.,1999), we know very little about its nature or its fundamental properties. The so-called consensus cosmological model, $\Lambda$CDM (Carroll et al. 1992; Sahni and Starobinski 2000), based on the well known cosmological constant, provides the current best fit to most of the cosmological data (e.g. Komatsu et al. 2011). However, it is well known that this model is also affected by serious theoretical shortcomings, such as the so-called "coincidence problem" and the "tuning" problem (Amendola and Tsujikawa, 2010). On the other hand, it is not currently possible, even using the latest combined results (SNe Ia, CMB, BAO and Cluster distribution) derived from ground and space observation surveys, to determine whether a cosmological constant or a more general constant field, a dynamical fluid field, i.e. a specific field which evolves with time, or a modification of general relativity, is the correct interpretation of dark energy. In principle, there is no a valid theoretical reason for considering the dark energy density a given constant as the universe evolves, and simple models for dark energy evolution can already be proposed and investigated with currently accessible cosmological probes (Weinberg et al. 2012).
A given fundamental component of the universe can be described, by its equation of state (in short: EoS), defined as the ratio of its pressure $p_X$ to its density, $\rho_X$: $w(a)=p_X(a) / \rho_X(a) ~ (c^2=1)$ where: $a = 1/(1 + z)$ is the scale factor of the universe and $z$ is the redshift (see for instance: Amendola and Tsujikawa, 2010). For matter, $w(a) = w = 0$, for radiation, $w(a) = w = 1/3$, and of course for dark energy this EoS parameter is currently unknown. If one assumes a constant EoS, wCDM model, the current observations place the constraints: $w = -0.990 \pm  0.085$  at $95\%$ confidence level (Burenin and Vikhlinin, 2012) which are consistent with the cosmological constant, $\Lambda$.
On the other hand, a significant effort is currently devoted to devise a physically adequate expression for an evolving dark energy EoS: the evolution of $w(z)$ can be reconstructed and verified from different observational data using either parametric or non-parametric methods. The first ones are based on an assumed mathematical form of $w(a)$ with some free parameters; whereas the second ones are based on a binning process without any assumed mathematical expression for $w(z)$. A common non-parametric approach is to bin $w(z)$ in $z$, or the scale factor $a$, and fit the bin amplitudes to real data. This process assumes that $w(z)$ is constant within each bin, while the neighboring bins are treated as independent. However, it seems rather unphysical and arbitrary to assume a perfect correlation of $w(z)$ within a given bin, while having no correlation between different bins. Here we consider a parameteric approach to an evolving dark energy, starting from the most popular form of this class of models, the so-called Chevallier-Polarski-Linder (CPL) model (Chevallier and Polarski 2001; Linder 2003), which is the common preferred model due to its simple formulation and adopted as reference scenario by many authors (see for instance, Salzano et al., 2013).
This parameterization defines the dark energy EoS as a two-parameter model: $w(a) = w_0 + (1-a)w_a$ or equivalently: $w(z) = w_0 + z/(1+z) w_a$, where $w_0$ is the value of EoS at the present time, and $w_0 + w_a$ is its asymptotic value at $a = 0$, i.e. at redshift $z \rightarrow \infty$.
Various attempts to generalize this expression have been made, pointing out the intrinsic difficulty to constrain the dynamic parameter $w_a$ with current datasets that are limited in their redshift range (such as SNe Ia, Salzano et al.,2013). In fact, when the EoS is allowed to evolve with the expansion, $waCDM$ with CPL model, the current observational constraints are much weaker: $w_0= -1.105 \pm  0.259; w_a=  0.388 \pm  0.936$ (at $95\%$ confidence level) (Burenin and Vikhlinin, 2012).  Among many different attempts to generalize the CPL model (Barai et al. 2004; Wetterich 2004; Choudhury and Padmanabhan 2005; Gong and Zhang 2005; Jassal et al. 2005; Lee 2005; Upadhye et al. 2005; Linder 2006; Lazkoz et al. 2010) a more recent interesting attempt is given by Ma and Zhang, 2011. In their work the authors point out a serious limit in the CPL model. In fact, the CPL model only explores the past expansion history of the Universe properly, but cannot describe the future evolution ($z < 0$) due to the fact that $\mid w(z) \mid$ grows increasingly and finally encounters a clear divergence as $z \rightarrow -1$.  To overcome this un-physically feature of the CPL model, the authors propose a novel parameterization forms of $w(z)$, maintaining the advantages of the CPL model, but extending its applicability in the future, avoiding its natural divergence.

We recall, as another example of early parameterized EoS models, that the three-parameter model by Wetterich, is of the form, (Wetterich, 2004):

\begin{equation}
 w(z)= \frac{w_0}{(1+b~ln(1+z))^2}
\end{equation}

where $b$ is the "bending parameter" which characterizes the redshift where there is a transition from a quite constant EoS to a different (dynamical) behavior, given by:

\begin{equation}
 b= \frac{3 w_0}{ln \left(\frac{1-\Omega_{DE}}{\Omega_{DE}}\right) + ln \left(\frac{1-\Omega_m}{\Omega_m}\right)}
\end{equation}

\section{Ma and Zhang parametric model of dynamical dark energy}

The leading proposal of Ma and Zhang parametric model for the EOS of dynamic dark energy is of the form (Ma and Zhang, 2011):

\begin{equation}
    w(z) = w_0 + w_a (ln(2 + z)/(1+z) - ln 2)
\end{equation}

where: $w_0$ and: $w_a$ have the same role as in the CPL model. It is clear that this new parameterization retains all the advantages of the original CPL model, i.e. it is sufficiently simple and useful to provide reliable predictions and sufficiently sophisticated to be able to accommodate the observational data. The idea of using a logarithmic function appears as the most natural choice of a good generalization of the original CPL formulation. In fact, the last term in the bracket is kept for maintaining the most important constraint for the static term $w_0$ in order to be the current observed value of $w(z)$. But the most important feature of this model is its bounded behavior for both high redshifts and negative redshifts. In fact as the author point out: "Thanks to the logarithm form in the parameterization, a finite value for $w(z)$ can be ensured, via the application of the L'Hospital's rule, in both limiting cases, $z \rightarrow \infty$ and $z \rightarrow -1$. This is the reason why we introduce a logarithm form in the new parameterization". The limiting cases for $w(z)$ are now:

$w(z)=$
$\begin{cases}
 w_0 & \text{for } z=0 \\
 w_0 - w_a ln 2 & \text{for } z \rightarrow \infty \\
 w_0 + w_a (1 - ln 2) & \text{for } z \rightarrow -1  
\end{cases}$

With this formulation (in short: MZp-model) we overcome the principal limit of CPL model, i.e. the divergence for $z \rightarrow -1$, in the future. Nevertheless, even this extended model is limited to probe an arbitrary $z$ value for the future evolution of EoS due to its shifted divergence for $z \rightarrow -2$. In order to validate the new parameterization, the authors explore the dynamical evolution of dark energy using both the CPL and MZp models. It is well-known from the observations that dark energy drives the cosmic acceleration only at recent times ($z \sim 0.3-0.5$), with a EoS around $-1$ in the recent epoch; whereas at the early times we can accept more possibilities for the EoS behavior. To probe the dynamics of dark energy, we start from the Hubble expansion rate through the equation:

\begin{equation}
\begin{gathered}
H^2(z)=\left(\frac{\dot a}{a}\right)^2=H_0 ^2 \left[\Omega_r(1+z)^4+\Omega_m(1+z)^3+\Omega_k(1+z)^2+ \right. \\
\Omega_{DE}~ exp \left(3 \int_{0}^{z} dz'(1+w(z'))/(1+z') \right) \left. \right]
\end{gathered}
\end{equation}

Requiring the consistency of this equation at z = 0, the present epoch, that is: $H(z = 0) = H_0$, gives: $\Omega_m+ \Omega_r+ \Omega_k+ \Omega_{DE} = 1$.
In order to simplify the constraint computations, the authors assume a flat universe, $\Omega_k = 0$, obtaining the following simpler expression for $H(z)$:

\begin{equation}
\begin{gathered}
H^2(z)=H_0 ^2 \left[\Omega_r(1+z)^4+\Omega_m(1+z)^3+(1- \Omega_r - \Omega_m) \right.\\ 
exp \left(3 \int_{0}^{z} dz'(1+w(z'))/(1+z') \right) \left. \right]
\end{gathered}
\end{equation}

where: $ \Omega_r = \Omega_\gamma(1+0.2271N_{eff})$, with: $ \Omega_\gamma = 2.469×10^{-5} h ^{-2}$ and: $h=H_0/100 ~ km ~ s^{-1} ~ Mpc^{-1}$. $N_{eff}$, is the effective number of neutrino species, here assumed equal to $3.04$ (Komatsu, 2011).

For constraining $w(z)$, with different parameterizations, it is common to use the current observational data from the type Ia Supernovae (SNe Ia), the baryon acoustic oscillations (BAO), and the cosmic microwave background (CMB). The model parameters considered in the best-fit process is a four-vector: $\mathbf{\theta}=(\Omega_m, w_0, w_a, h)^T$.
 
The best-fit solutions found by Ma and Zhang for CPL and logarithm parameterizations are ($1 \sigma$ confidence level):
\begin{equation}
(\Omega_m=0.279^{+0.032}_{-0.028} ~ w_0=-1.066^{+0.267}_{-0.232} ~ w_a=0.261^{+0.904}_{-1.585} ~ h=0.699+^{+0.029}_{-0.034})^T ~ \chi^2_{min} = 544.186
\end{equation}
\begin{equation}
(\Omega_m=0.280^{+0.032}_{-0.028} ~ w_0=-1.067^{+0.234}_{-0.155} ~ w_a=-1.049^{+5.706}_{-0.896} ~ h=0.697+^{+0.031}_{-0.026})^T ~ \chi^2_{min} = 544.081,
\end{equation}
respectively.

Moreover, the authors found that, considering the whole evolutionary history of $w(z)$ from past to future via the fitting results, the proposed logarithm parameterization is more tightly constrained by the data (see: Ma and Zhang, 2011). However, as said, even this extended model is limited to probe an arbitrary z value for the future evolution of EoS, due to its shifted, but persistent, divergence for: $z \rightarrow -2$.

\section{A simple generalization model of dynamical dark energy}

As a simple generalization of the proposed MZp-model, which is not restricted to a fixed value of z in the future due to a related divergence, we suggest to use a more general parameterized version of the following form:

\begin{equation}
    w(z) = w_0 + w_a (ln(\xi+ 1+ z)/(\xi+z) - ln (\xi+1)/\xi),
\end{equation}

with $\xi$ a given positive parameter. In the particular case of $\xi=1$ we obtain the above MZp-model. As in the MZp-model, in this general model, a finite value for $w(z)$ can be ensured, via the application of the L'Hospital's rule, in both limiting cases, $z \rightarrow \infty$ and $z \rightarrow -\xi$, and moreover it is also allowed to explore more far values in the future up to $z \rightarrow -(\xi+1)$ where there is a natural divergence. The limiting cases for $w(z)$ are now:

$w(z)=$
$\begin{cases}
 w_0 & \text{for } z=0 \\
 w_0 - w_a ln(\xi+1)/\xi & \text{for } z \rightarrow \infty \\
 w_0 + w_a (1-ln(\xi+1)/\xi) & \text{for } z \rightarrow -\xi  
\end{cases}$

Of course, there is again a divergence in this model, but for an arbitrary longer future value: $z \rightarrow -(\xi+1)$. In fact, the main advantage of this simple generalized model (in short: $\xi$MZp-model), with respect to MZp-model, is the possibility to explore more far regions in the future of the EoS of dark energy ($z < -2$). This model retains all the good features of the MZp-model, and in particular finite limiting values and the consistency for $z=0$, the present epoch. 

However, in order to test the performances of this parameterization on a wider evolutionary history of $w(z)$ from past to future, it is necessary to compute new fitting results for vector $\mathbf{\theta}$ with this extended formulation of $w(z)$, and to verify if we obtain more tightly constraints by the data. This would be the task of a next work. Indeed, the main aim of the present work is to suggest new and more generalized parameterizations for the EoS of dark energy, covering both the past and the future of the cosmic history, consistently with the current observational data.
In the process of best-fitting with observational data to obtain solutions as eq.(4) or eq.(5), for $H(z)$, the dependence on the EoS of dark energy is contained in the integral term of equations, eq.(2) or eq.(3). For that reason it is useful to give an analytical expression of these integrals related to different EoS parameterization models. For CPL model, the computation of the integral gives the following expression:

\begin{equation}
 \int_{0}^{z} dz'(1+w(z'))/(1+z') = (1+w_0+w_a) ln (1+z) + w_a/(1+z) -w_a
\end{equation}

Note that the integral term in the expression for the evaluation of Hubble parameter, $H(z)$, is undefined for: $z\leq -1$.

For MZp-model, eq.(3),the computation of the integral term gives the following expression:

\begin{equation}
\begin{gathered}
 \int_{0}^{z} dz'(1+w(z'))/(1+z') = (1+w_0-w_a ~ln(2))~ ln(1+z)\\ 
+ w_a \left(ln(1+z)-\frac{log(2+z)(2+z)}{(1+z)}+2~ln(2)\right)
\end{gathered}
\end{equation}

whereas, for general $\xi$MZp-model, the computation of the integral gives the following expression:

\begin{equation}
\begin{gathered}
 \int_{0}^{z} dz'(1+w(z'))/(1+z') = (1+w_0-w_a ~ln(\xi+1)/\xi )~ ln(1+z) + w_a \left[ 1/(\xi-1) \cdot \right.\\
  \cdot \left( L_2(1-x)+ln(\xi)ln \left(\frac{x-\xi}  {\xi} \right)-L_2(1- x/\xi)\right)_{(\xi+1)}^{(\xi+1+z)} \left. \right]
\end{gathered}
\end{equation}

where: $\xi>1$ and: $L_2(1-x)=-\int_{0}^{1-x} dt \left( \frac{ ln(1-t)}{t} \right)$ is the Euclidean dilogarithm function. As examples: $L_2(-1)=-\pi^2/12$, and: $L_2(0)=0$.

 \section{A new divergence-free generalization model of dynamical dark energy}

The proposed $\xi$MZp-model, eq.(8), retains the unwanted feature of an intrinsic divergence in an arbitrary but definite far future, limiting again the applicability of this parameterized model for EoS of dynamical dark energy. Moreover, we have to choose a value for the limit $\xi$ parameter where the model diverges and prevent its use for $z \rightarrow -(\xi+1)$. The main aim of a next generalized formulation is to obtain a divergence-free model of dynamical dark energy, maintaining the basic structure of two-parameters CPL-model, in order to explore arbitrary regions in the future without encountering unwanted divergences.
Our proposal for a divergence-free parametric model (in short: DFp-model) for the EoS of dynamic dark energy is of the form:

\begin{equation}
w(z) = w_0 + w_a \left(\frac{ln\sqrt{1+z^2} -ln\sqrt 2}{1+z} + ln\sqrt 2 \right)
\end{equation}

As in previous models, in this new general formulation, a finite value for $w(z)$ can be ensured, via the application of the L'Hospital's rule, in both limiting cases, $z \rightarrow \infty$ and $z \rightarrow -1$, and moreover it is also allowed to explore any far value in the future without encountering divergences. The additional term: $+ln\sqrt 2$, ensures that for: $z=0$, we have: $w(z)=w_0$; whereas the term: $-ln\sqrt 2$ ensures the treatment in the limit case $z\rightarrow -1$. The limiting cases for $w(z)$ are now:

$w(z)=$
$\begin{cases}
 w_0 & \text{for } z=0 \\
 w_0 + w_a ln\sqrt 2  & \text{for } z \rightarrow \infty \\
 w_0 - w_a (1/2 + ln\sqrt 2) & \text{for } z \rightarrow -1  
\end{cases}$

and, more important, it is now possible to explore every negative value of $z$ for the future evolution of $w(z)$ without encountering any divergence.
The DFp-model, eq.(12), retains all the good features of previous models and, in particular, finite limiting values other than the consistency for $z=0$, the present epoch. However, as already said, in order to test the performances of this divergence-free parameterization on a wider evolutionary history of $w(z)$ from past to future, it is necessary to compute new fitting results for vector $\mathbf{\theta}$ with this extended formulation of $w(z)$, and to verify if we obtain more tightly constraints by the current data. This would be the task of a future work.

For DFp-model, the computation of the integral term containing the EoS of dark energy, gives the following expression:

\begin{equation}
\begin{gathered}
$
 $\int_{0}^{z} dz'(1+w(z'))/(1+z')
 = \left(1+w_0+w_a ~ln{\sqrt{(2)}} \right)ln(1+z) -w_a ~ ln{\sqrt{(2)}} \\
\left(1- \frac{1}{(1+z)}\right) + w_a \left[- \frac{1}{2}\frac{1}{(1+z)} \left( ln \left( -2z-2+\sqrt{(2)} arctanh(\frac{1}{2}\sqrt{(2)})z  \right. \right. \right.\\
+\sqrt{(2)} arctanh(\frac{1}{2}\sqrt{(2)})+(1+z^2)^{3/2}+\sqrt{(2)}
arctanh(\frac{1}{2}(z-1)\sqrt{(2)}/\sqrt{(1+z^2)})+ \\
\sqrt{(1+z^2)}-\sqrt{(1+z^2)}z^2+2 ~ln(-z+\sqrt{(1+z^2)})+2~ln(-z+\sqrt{(1+z^2)})z \\
+\sqrt{(2)} arctanh(\frac{1}{2}(z-1)\sqrt{(2)}/\sqrt{(1+z^2)})z \left. \right) \left. \right) \left. \right],$
$
\end{gathered}
\end{equation}

for: $z>0$, and:
\begin{equation}
\begin{gathered}
$
$\int_{0}^{z} dz'(1+w(z'))/(1+z')
 = \left(1+w_0+w_a ~ ln{ \sqrt{(2)} } \right)ln(1+z) -w_a ~ ln{\sqrt{(2)}} \\
\left(1- \frac{1}{(1+z)}\right) + w_a \left[- \frac{1}{2}\frac{1}{(1+z)} \left(ln \left( (1+z^2)^{3/2}+\sqrt{(2)}arctanh\left(\frac{1}{2}\frac{(z-1)\sqrt{(2)}}{\sqrt{(1+z^2)}} \right) \right. \right. \right. \\
+\sqrt{(1+z^2)}-\sqrt{(1+z^2)}z^2-2~ln(z+\sqrt{(1+z^2)})-2~ln(z+\sqrt{(1+z^2)})z+ \\
\sqrt{(2)} arctanh\left(\frac{1}{2}\frac{(z-1)\sqrt{(2)}}{\sqrt{(1+z^2)}} \right)z -2z-2+\sqrt{(2)}arctanh(\frac{1}{2}\sqrt{(2)})z \\
+\sqrt{(2)}arctanh(\frac{1}{2}\sqrt{(2)})
\left. \right) \left. \right) \left. \right],$
$
\end{gathered}
\end{equation}

for: $-1<z<0$.  

As in previous cases, the integral term in the expression for the evaluation of the Hubble parameter, $H(z)$, is undefined for: $z \le -1$.

Despite the more complex mathematical expression of integral terms, eq.(13) and eq.(14), the potential advantages of using the general parameterization, eq.(12), for both $z$ positive and negative without encountering divergences for the EoS of dark energy, justify the additional effort needed to handle and to compute the above terms. 

\section{Conclusions}

The main aim of the present work is to suggest new and more generalized parameterizations for the EoS of dark energy, maintaining the basic structure of two-parameters CPL-model, but covering both the past and the future of the cosmic history, without divergences and consistently with the current observational data. We proposed two generalizations, starting from the extended MZp-model by Ma and Zhang, 2011, the $\xi$MZp-model and the DFp-model. The former extends its validity in the future, through the introduction of a new parameter $\xi$, shifting the natural divergence of the model up to: $z=-(\xi+1)$; the latter, in principle more interesting, extends its validity for any values of negative $z$, in the future, being free of any unwanted divergence. Of course, the physical interest of these consistent mathematical formulations, at least for $z=0$, needs to be verified on a wider evolutionary history of $w(z)$ from past to future, computing new fitting results for vector $\mathbf{\theta}$ with current observational data using these new parameterizations, and to verify if we obtain more tightly constraints with respect to older and more popular parametric models. The potential advantages of using these new formulations is their extended range of validity, mainly in the future, to determine possible future scenarios of the cosmic evolution.

\section{References}

Amendola, L. and Tsujikawa, S., 2010, Dark Energy-Theory and Observations, Cambridge University Press.

Barai, P., Das, T.K., Wiita, P.J., 2004, Astrophys. J. 613, L49.

Burenin R.A., Vikhlinin, A.A., 2012 arXiv:1202.2889.

Carroll, S.M., Press, W.H., Turner, E.L., ARA\&A 30 (1992) 499.

Chevallier M., Polarski D., 2001 Int. J. Mod. Phys., D10, 213.

Choudhury, T.R., Padmanabhan, T., 2005, Astron. Astrophys. 429, 807.

Gong, Y.G., Zhang, Y.Z., 2005, Phys. Rev. D 72, 043518.

Jassal, H.K., Bagla, J.S., Padmanabhan, T., 2005, Mon. Not. R. Astron. Soc. 356, L11.

Komatsu E. et al. 2011, Astrophys. J. Suppl. 192, 18.
 
Lazkoz, R., Salzano, V., Sendra, I., 2010, Phys. Lett. B 694, 198.

Lee, S., 2005, Phys. Rev. D 71, 123528.

Linder E. V., 2003 Phys. Rev. Lett., 90, 091301.

Linder, E.V., 2006, Phys. Rev. D 73, 063010.

Ma Jing-Zhe and Zhang Xin, 2011, arXiv:1102.2671v2, Phys. Lett. B 699, 233.

Perlmutter S. et al., 1999, ApJ, 517, 565.

Riess A. G. et al., 1998, AJ, 116, 1009.

Sahni, V., Starobinski, A., 2000, Int. J. Mod. Phys. D 9  373.

Salzano, V. et al., 2013 arXiv:1307.0820v1.

Upadhye, A., Ishak, M., Steinhardt, P.J., 2005, Phys. Rev. D 72, 063501.

Weinberg, D.H., Mortonson, M.J., Eisenstein, D.J., Hirata, C., Riess, A.G., Rozo, E., 2012, arXiv:1201.2434.

Wetterich, C., 2004, Phys. Lett. B 594, 17.

\end{document}